\documentclass[final,english]{bullsrsl}

\usepackage[latin1]{inputenc}
\usepackage[T1]{fontenc}

\usepackage{natbib} 
\usepackage{graphicx}
\usepackage{xspace}

\newcommand{\tninty}{{$T_{\rm 90}$}\xspace}

\begin{document}
\title{Investigating high redshift short GRBs: signatures of collapsars?}

\author[affil={1,2}, corresponding]{Dimple}{}
\author[affil={1}]{Kuntal}{Misra}
\author[affil={2}]{Lallan}{Yadav}

\affiliation[1]{Aryabhatta Research Institute of Observational Sciences (ARIES), Manora Peak, Nainital-263002, India.}
\affiliation[2]{Department of Physics, Deen Dayal Upadhyaya Gorakhpur University, Gorakhpur-273009, India.}
\correspondance{dimplepanchal96@gmail.com}

\date{}
\maketitle

\begin{abstract}
The conventional classification of Gamma-Ray Bursts (GRBs) as short or long bursts based on their duration is widely accepted as arising from different progenitor sources identified as compact object mergers and collapsars, respectively. However, recent observational shreds of evidence challenged this view, with signatures of collapsars in short GRBs and mergers in long GRBs. We conduct a comparative analysis of the characteristics of short and long GRBs, both at low and high redshifts, taking into account the locations and environments of their host galaxies. Our analysis suggests that some short GRBs at higher redshifts exhibit features similar to long GRBs, indicating a possible collapsar origin. Further investigation, utilizing multi-messenger observations, could provide a resolution to this issue. 

\end{abstract}


\section{Introduction}
\label{introduction}
The bimodality in the duration distribution of GRBs suggested two broad classes as long and short determined by the \tninty duration (the time interval of integrated counts between 5\% to 95\%) with a boundary at 2 seconds \citep{Kouveliotou1993}. The long-duration GRBs, with \tninty  $> 2$ sec, were postulated to stem from the death of the massive stars \citep{Woosley1993,Mszros2006}, while the short-duration, with \tninty  $ \leq 2$ sec, were believed to stem from mergers involving compact objects \citep{Paczynski1986, Meszaros1992}. Multi-wavelength observations of GRBs provided evidence in support of these predictions, as several long GRBs have been discovered to be linked with Type Ic supernovae  \citep{Woosley_1993, MacFadyen1999, Hjorth2003, Woosley2006, Cano2017}, and the origin of short-duration GRBs from compact object mergers is supported by the coincident discovery of GW170817/GRB~170817A and the associated kilonova AT 2017gfo \citep{Abbott2017, Valenti_2017}.

The established theory of long and short GRBs origin became questionable after the detection of a supernova bump associated with a short-duration GRB in August 2020, GRB~200826A \citep{zhang_2021,Ahumada2021,Rossi2022}, and a long-duration GRB identified with a kilonova bump in December 2021, GRB~211211A \citep{Rastinejad2022, Troja2022, Yang2022}. Recently, another long GRB~230307A is found to be associated with a Kilonova \citep{Levan_2023}. The dichotomous separation of GRBs based on duration has been questioned time and again \citep{Fynbo_2006, zhang2009}. In the past, numerous efforts have been undertaken to create novel classification systems utilizing criteria distinct from the traditional \tninty classification. For instance, \citet{Zhang_2006} categorized GRBs into Type I (arising from compact binary mergers) and Type II (arising from death of massive stars) groups. \citet{Bromberg2013} employed a classification based on collapsar and non-collapsar probabilities. Additionally, \citet{Mineav_2020} utilized the $E_{\rm \gamma, iso}$ - $E_{\rm p,i}$ correlation to divide GRBs into two distinct classes. 
Additional parameters such as hardness ratio, spectral lag, and variability time scales in light curves were identified to differentiate between distinct progenitors of GRBs \citep{Fishman1995,Bernardini2015,McInnes2018}. 
However, these parameters have a significant overlap for the two classes, making classification challenging as argued by \cite{Dimple2022}. Distinction between long and short GRBs still remains challenging.

The redshift distributions of long and short GRBs provide essential clues about their progenitor systems \citep{Guetta2005, Berger2007, Ghirlanda2009AA, Avanzo2015}. The median redshifts observed for long and short GRBs strongly suggest that these events originate from different types of progenitors. The higher redshift of long GRBs agrees with the  predictions of rapidly evolving massive star progenitors. In contrast, the lower redshift of short GRBs matches with the longer timescales of compact object mergers \citep{Berger2013}. However, a fraction of short GRBs is found to lie at high redshifts contradicting their proposed progenitors. Given the overlap observed between the two classes, it is reasonable to investigate the properties of short and long GRBs at both low and high redshifts.
   
Our recent work on the comparison of low and high redshift short GRBs suggested that they could be arising from different progenitor systems \citep{Dimple2022b}. We expand on this work with an updated sample and identify their position on the Amati plot \citep{Amati_2006} as well as compare their offset (The distance between the burst and the centre of its host galaxy) and number density (the density of the ambient medium) distributions. Both offset and number density aid in inferring the GRB progenitors. The description of the sample and comparison of the short GRB properties (Amati correlation, offset, number density, non-collapsar probability; $F_{\rm nc}$ and star formation rate; SFR of GRB hosts) are given in Section \ref{comparison}. The key findings are summarised in Section \ref{summary}.

\section{Are short GRBs at high redshift similar to long GRBs?}
\label{comparison}
The redshift distribution (the left panel of Figure \ref{z_amati}) shows that short GRBs are found at significantly lower redshifts as compared to long GRBs. The sample is compiled from Jochen Greiner's long GRB table\footnote{\url{https://www.mpe.mpg.de/~jcg/grbgen.html}} with known spectroscopic redshifts and is complete upto April 2023. We estimated a median redshift of short GRBs to be $z = 0.51$, much lower than the median redshift of long GRBs at $z = 1.65$, consistent with the expectations for their progenitors. However, a notable proportion of short GRBs is observed to occur at higher redshifts ($z > 0.7$). Out of these, GRBs~200826A \citep[$z=0.7481$;][]{Rossi2022} and 090426 \citep[$z=2.609$;][]{Antonelli_2009}, have been suggested to originate from collapsars rather than compact object mergers. 
It is crucial to investigate whether the progenitors of short GRBs at low and high redshifts differ and if the high redshift short GRB progenitors share similarities with those of long GRBs. As suggested by \citet{Berger2007} and  \citet{Bromberg2013} that the short GRBs detected at higher redshifts may represent a distinct population of GRBs, our investigation explores the similarities and dissimilarities between short GRBs lying at low and high redshifts and that with long GRBs.

Since, a short GRB has been observed to be associated with supernova at a redshift of 0.7481 (GRB~200826A) and around 43\% of short GRB lie at $z > 0.7$. Therefore, we have put a cut at $z = 0.7$. For this, we plot the Amati correlation and examine their offset and number density distributions, and the $F_{\rm nc}$ and SFR estimations of the hosts to investigate their progenitor systems.

\subsection{Amati correlation}
\begin{figure*}[!h]
    \bigskip
    \begin{minipage}{\linewidth}
    \includegraphics[width=0.49\columnwidth]{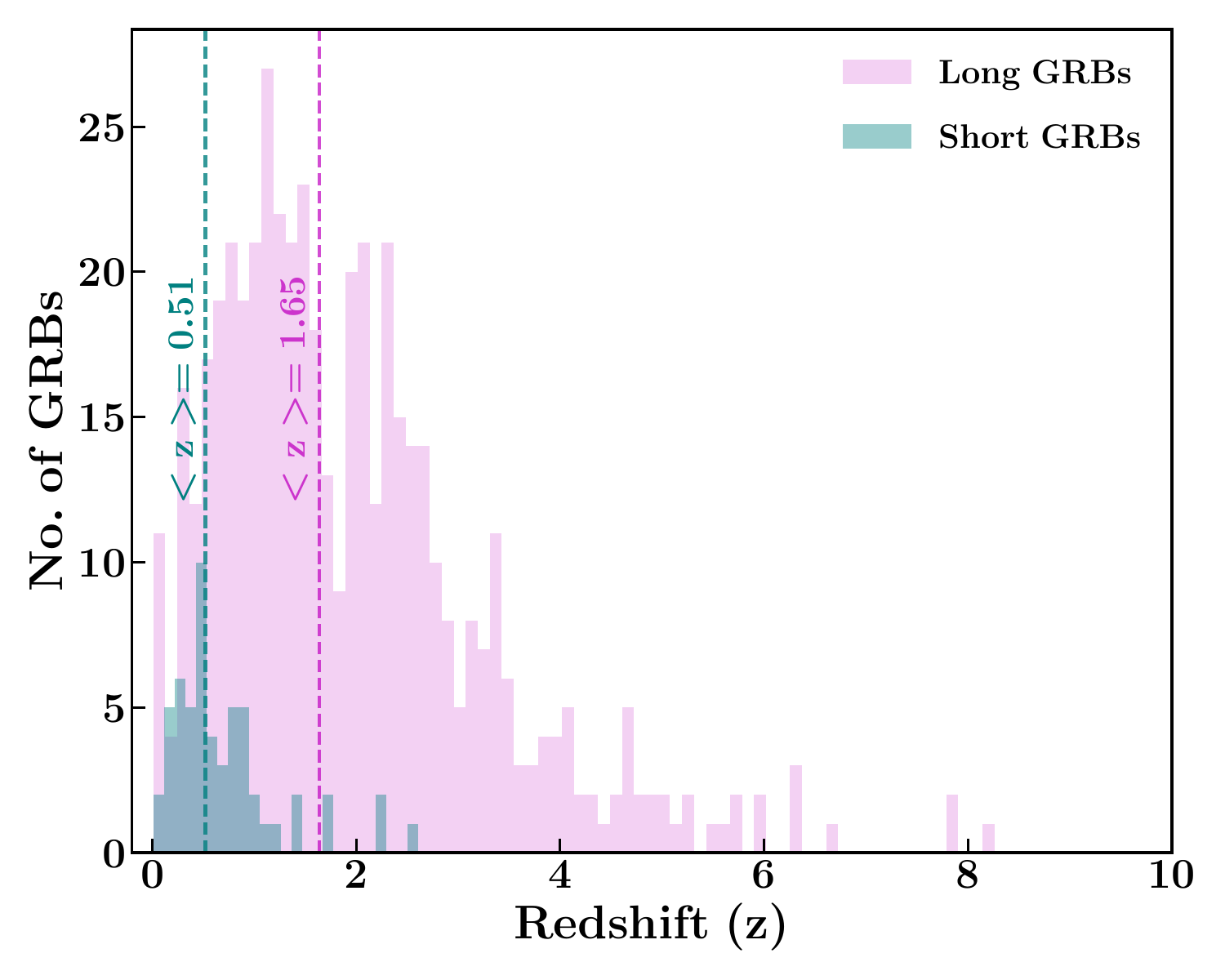}
    \includegraphics[width=0.49\columnwidth]{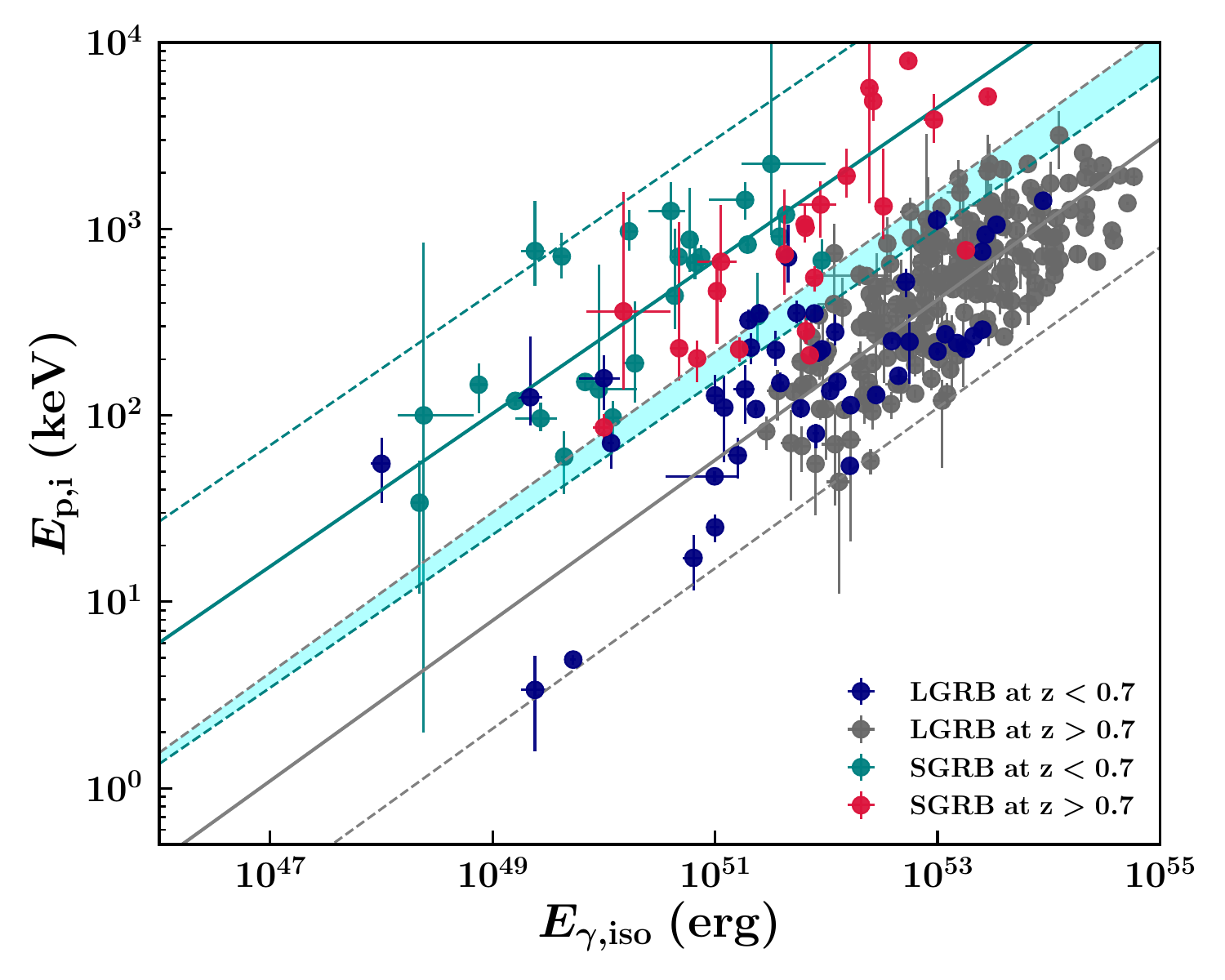}
    \caption{{\bf Left:} The redshift distributions of GRBs up to April 2023. Short GRBs occupy the lower end of the distribution, with a median redshift of 0.47.  In contrast, long GRBs span a wider range of redshifts, with a median of 1.68.
    {\bf Right:} Short and long GRBs in the Amati correlation plane. The solid blue and grey lines in the plot depict the best-fit correlation for long and short GRBs, respectively, while the dotted lines illustrate the corresponding $2-\sigma$ regions. The shaded region shows the overlapping $2-\sigma$ regions between long and short GRBs. It can be observed that at redshifts z $>0.7$, some short GRBs follow the long GRB track, while some fall within the overlapping $2-\sigma$ region.  Similarly, some long GRBs at lower redshifts follow the short GRB track}
    \label{z_amati}
      \end{minipage}
\end{figure*}

For a long time, correlations in prompt emission have been utilized to categorize GRBs. In the Amati correlation plane, which plots $E_{\rm \gamma, iso}$ against $E_{\rm p}$ (the peak energy in the source frame), two distinct classes of GRBs follow different tracks, and occupy different positions \citep{Amati_2002,Amati_2006}. To check the Amati correlation of GRBs lying at various redshifts, we took the isotropic and peak energy values from \cite{Mineav_2020} and divided the sample according to the redshift of the data with a divider at a redshift of 0.7. Short GRBs lying at redshift $> 0.7$ are located on the long GRB track, while some of them are in the overlapping $2\sigma$ correlation region (cyan shaded region) between short and long GRBs, as can be noticed from the right panel of Fig. \ref{z_amati}. Similarly, some of the long GRBs lying at lower redshifts are located on the short GRB track. The presence of short GRBs at high redshifts aligning with the long GRB track and low-redshift long GRBs falling on the short GRB track can be due to the selection effects or these GRBs originate from progenitors that differ from their identified class based on duration.

\subsection{Offset Distribution}
The offset of a GRB from its host galaxy can provide meaningful insight into its progenitor system. Binary compact stellar remnants are expected to merge far from their birth sites due to the kicks imparted to the neutron stars during supernova explosions \citep{Belczynski:2002ApJ,Belczynski:2006ApJ}. As a result, GRBs arising from these progenitors are expected to have large offsets from their host galaxies \citep{Berger2013}. However, in the case of massive star collapse, GRBs are expected to occur near their birth site, likely in an active star-forming region \citep{Fruchter2006}. As a result, GRBs arising from collapsars are likely to have smaller offsets from their host galaxies which has also been confirmed observationally \citep{Bloom_2002}. 

\begin{figure*}[!h]
    \bigskip
    \begin{minipage}{\linewidth}
    \includegraphics[width=0.49\columnwidth]{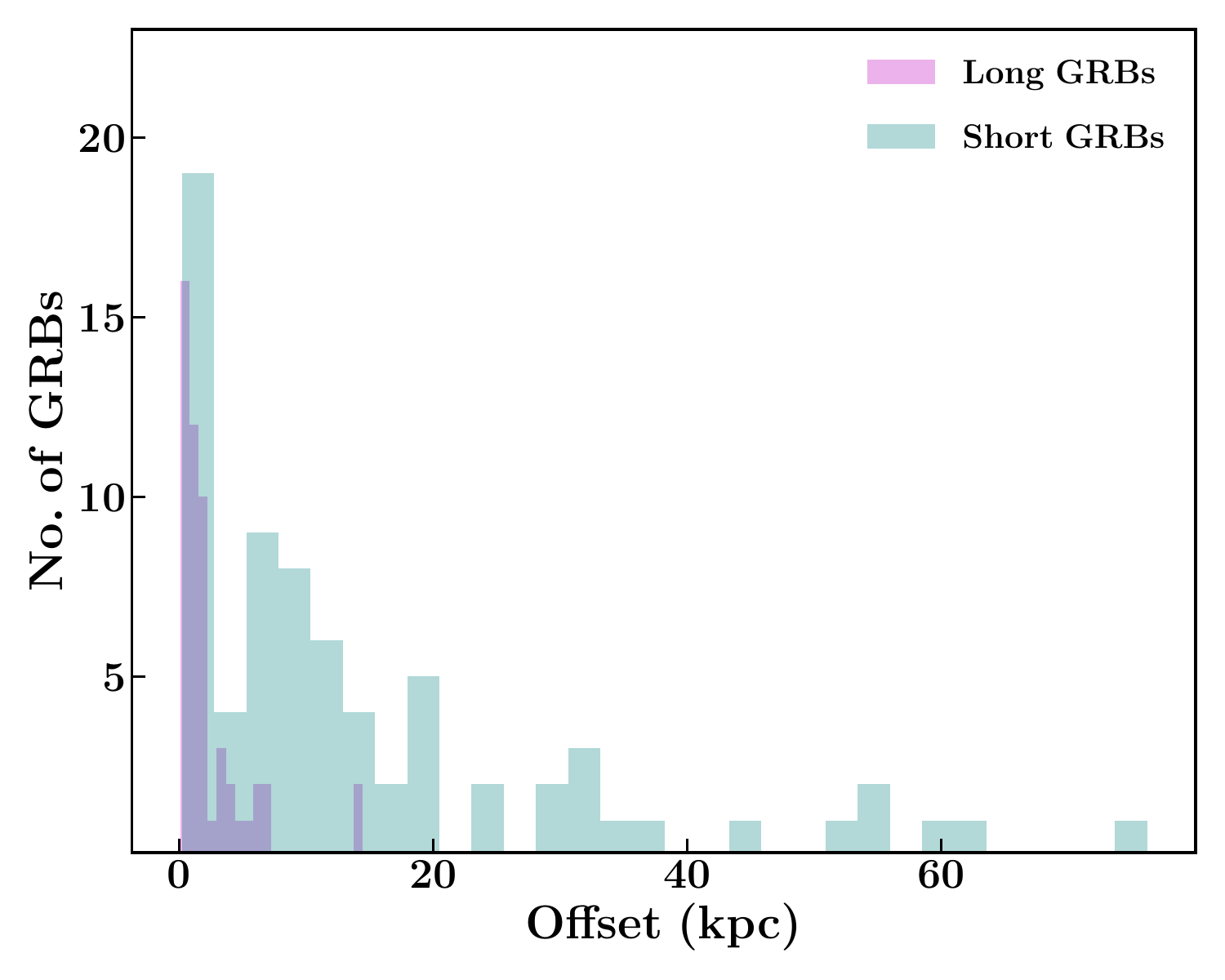}
    \includegraphics[width=0.49\columnwidth]{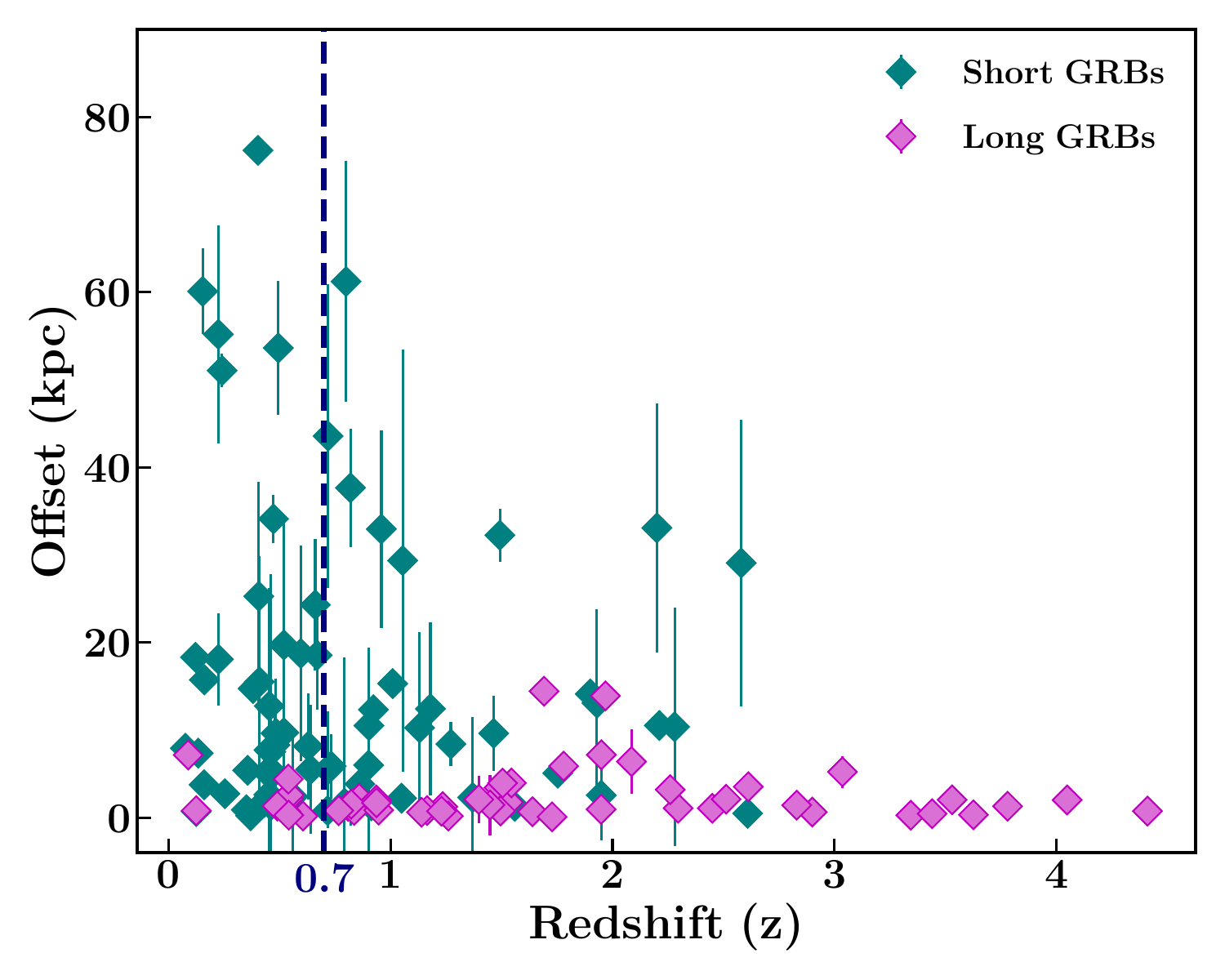}
    \caption{{\bf Left:} Offset distribution of  GRBs. Short GRBs display a median offset of approximately 9.62 kpc, which is roughly six times larger than the median offset observed for long GRBs (1.38 kpc).
    {\bf Right:} Variation of offsets with redshifts for GRBs with a cut at redshift = 0.7.}
      \label{offset_distribution}
        \end{minipage}
\end{figure*}

Figure \ref{offset_distribution} illustrates the offset distributions of long and short GRBs using data from \cite{Blanchard_2016} and \cite{Fong_2022}, respectively. From the left panel of the figure, we deduce that the projected offsets of short GRBs range from 0.23 to 76.19 kpc, with a median offset of approximately 9.62 kpc. This median offset is approximately six times larger than the median offset observed for long GRBs, which is approximately 1.38 kpc. The distribution of short GRBs aligns with the predictions of population synthesis models for compact object mergers, especially regarding the fraction of events displaying significant offsets. Conversely, long GRBs exhibit significantly lower offsets, consistent with the expectations of massive star progenitors exploding in close proximity to their birth site within the host galaxy.

The right panel of Figure \ref{offset_distribution} presents the variation of offset with the redshift of GRBs. It is apparent that short GRBs at higher redshifts tend to have slightly smaller offsets compared to those at lower redshifts, although they still remain noticeably larger than the offsets observed for long GRBs. While a few short GRBs display offsets similar to those of long GRBs, the overall offset distribution strongly supports the existence of two distinct progenitor populations for long and short GRBs. Further investigation of short GRBs with lower offset values would be of great interest to understand their unique characteristics and progenitor systems.

\subsection{GRB Environments: Number densities}
\begin{figure*}[!h]
    \bigskip
    \begin{minipage}{\linewidth}
    \includegraphics[width=0.49\columnwidth]{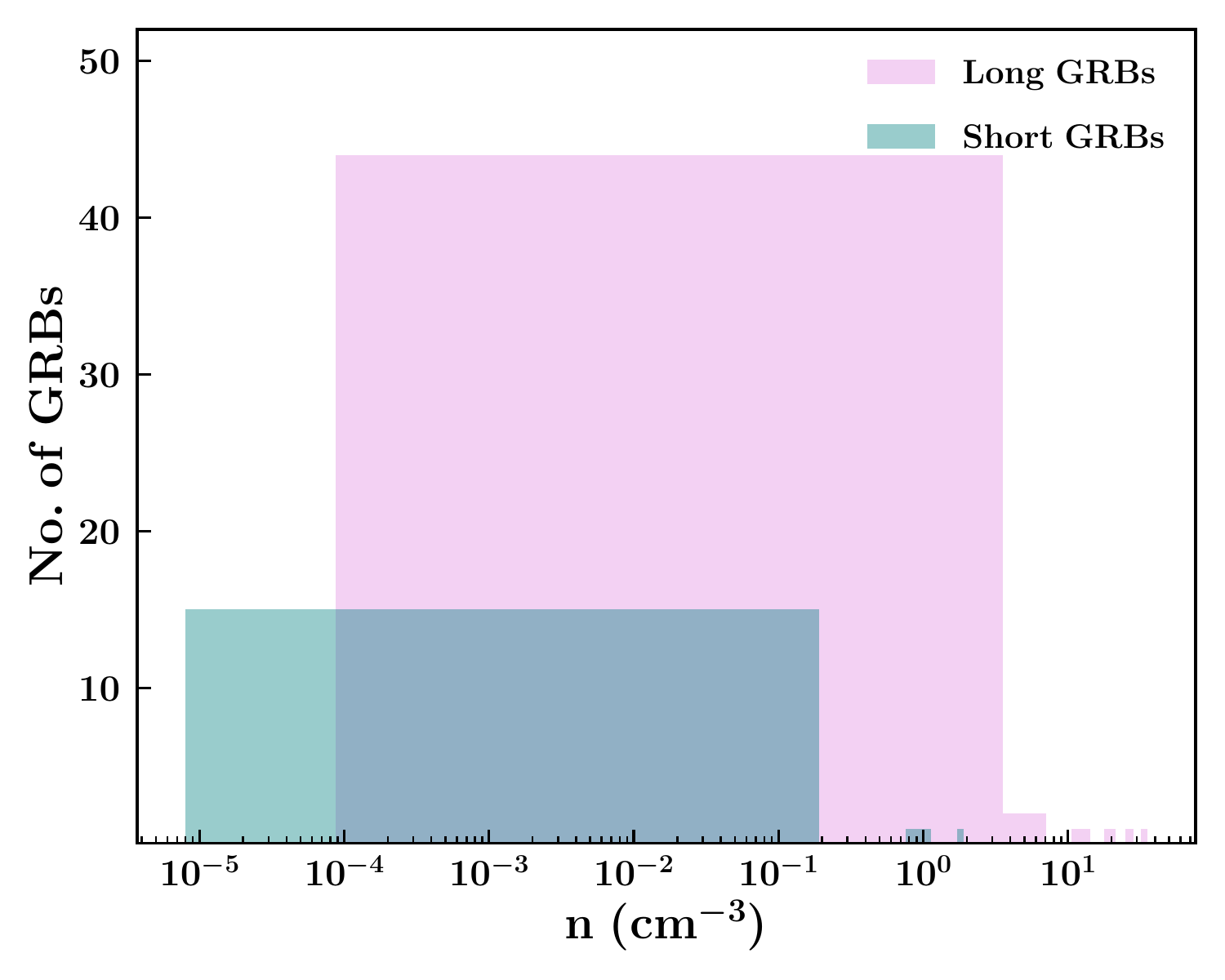}
    \includegraphics[width=0.49\columnwidth]{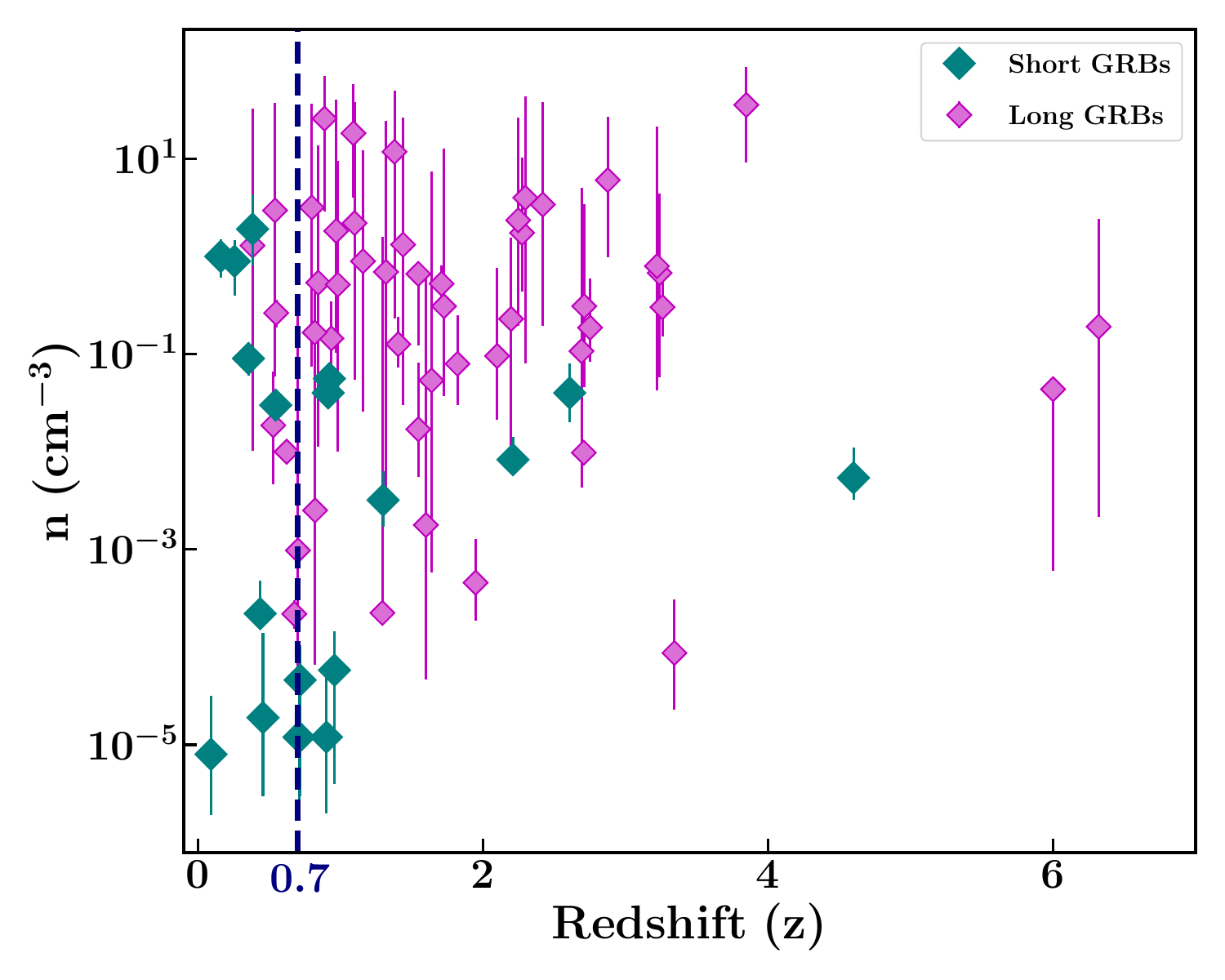}
    \caption{{\bf Left:} Number density distribution for long and short GRB environments. Median number densities are 0.315 $cm^{-3}$ for long GRBs and 0.0068 $cm^{-3}$ for short GRBs. 
    {\bf Right:} Variation of the number density of GRB environments with redshift with a cut at z = 0.7.}
\label{num_density}
  \end{minipage}
\end{figure*}

Both the progenitors (collapsars and mergers) also differ in the number densities of their environments. Long-duration GRBs are often linked to star-forming regions exhibiting high gas densities, indicating the presence of dense environments surrounding these events. Conversely, short-duration GRBs are found in low-density environments \citep{Connor_2020}, possibly due to natal kicks that propel them into the wider interstellar medium or even the intergalactic medium \citep{Berger2013}.
In order to examine the variation of number density for long and short GRBs, we utilized a sample of long GRBs from \cite{Chrimes_2022} and short GRBs from \cite{Fong_2015}. Both the studies have estimated the environmental number densities by fitting the  afterglow data with standard afterglow model (the forward-reverse shock model, \citealt{Sari_1998}). The results are illustrated in Figure \ref{num_density}. The median number densities for long and short GRBs are 0.315 $cm^{-3}$ and 0.0068 $cm^{-3}$, respectively. There is a substantial overlap observed in the number densities of long and short GRBs; however, the short GRB distribution is skewed more towards low number densities and long GRB towards high number densities. The right panel of Figure \ref{num_density} shows the variation of number densities with redshift. It can be seen that the short and long GRBs show similar trends with low redshift short GRBs typically favouring lower densities up to $10^{-5} cm^{-3}$.

\section{Discussions}
\label{summary}
To investigate whether long and short GRBs with known redshifts have similar progenitor systems, we compared various properties between them in detail. We located low and high redshift GRBs on the Amati plane. Overall, short and long lie on two different tracks, however, some of the GRBs show peculiarity and share a different track. Interestingly, these are high redshift short GRBs and low redshift long GRBs. Further, we found similarities between high-redshift short GRBs and long GRBs in terms of offsets and environmental densities with redshift. Also, the low values of $F_{\rm nc}$ and high values of SFR of short GRBs support the argument that high-redshift short GRBs could have progenitors similar to those of long GRBs \citep{Dimple_2023,Dichiara_2021,Bromberg2013}. However, this can also be an observational bias. More studies are needed to understand better the role of redshift in GRB classification and the reliability of inferring progenitors based on empirical correlations. Future multimessenger observations, including optical and near-infrared observations, will be instrumental in detecting bumps in GRBs' light curves, which can help in identifying supernovae/kilonovae associated with collapsars/compact binary mergers. These observations, accompanied by gravitational wave detections, will provide additional information on the properties of the progenitor. In addition, machine learning algorithms play a crucial role in clustering the GRBs based on fine structures present in their light curves. \cite{Christian2020} and \cite{Steinhardt2023} have used t-distributed Stochastic Neighbor Embedding (tSNE) and Uniform Manifold Approximation and Projection (UMAP), respectively, to identify clustering in the population of GRBs. \cite{Luo_2022} also made use of supervised machine learning techniques to distinguish between different progenitor systems of GRBs. However, these studies focused more on the general classification of GRBs based on their light curves and did not survey any specific subpopulation, such as that of the KN-associated GRBs. Recently, \cite{Keneth_2023} has used tSNE to investigate the extended emission GRBs. We employ the Principal Component Analysis (PCA) in conjunction with tSNE and UMAP to the light curves to cluster the GRB population \citep{Dimple_2023}. We will investigate the short GRBs at high redshifts using machine learning algorithms in near future.

\begin{acknowledgments}
The authors wish to thank Prof. K. G. Arun and Prof. L. Resmi for their useful discussions.
\end{acknowledgments}

\begin{furtherinformation}

\begin{orcids}
\orcid{0000-0001-9868-9042}{Dimple} {}
\orcid{0000-0003-1637-267X}{Kuntal}{Misra}
\end{orcids}

\begin{authorcontributions}
Toward the completion of this work, all authors have made significant contributions.
\end{authorcontributions}

\begin{conflictsofinterest}
The authors declare no conflict of interest.
\end{conflictsofinterest}

\end{furtherinformation}

\bibliographystyle{bullsrsl-en.bst}

\bibliography{BINA}

\end{document}